\documentclass{jpp}
\usepackage{graphicx}

\usepackage[utf8]{inputenc}
\usepackage[T1]{fontenc}
\usepackage{amsmath}

\title{Direct prediction of saturated neoclassical tearing modes in slab using an equilibrium approach}

\author{E. Balkovic\aff{1}, J. Loizu\aff{1}, J. P. Graves\aff{1}, Y.-M. Huang\aff{2}, C. Smiet\aff{1}}

\affiliation{\aff{1} \'Ecole Polytechnique F\'ed\'erale de Lausanne, Swiss Plasma Center, CH-1015 Lausanne, Switzerland \\ \aff{2} Princeton University, Princeton NJ 08543, USA}

\newcommand{\deriv}[2]{\frac{\partial #1}{\partial #2}}
\newcommand{\paran}[1]{\left( #1 \right)}
\newcommand{\tensor}[1]{\mathsfbi{#1}}
\newcommand{\vect}[1]{\boldsymbol{#1}}
\newcommand{\vecdot}[0]{\boldsymbol\cdot}

\begin{document}

\maketitle
\date{}
\begin{abstract}

We demonstrate for the first time that the nonlinear saturation of neoclassical tearing modes (NTMs) can be found directly using a variational principle based on Taylor relaxation, without needing to simulate the intermediate, resistivity-dependent dynamics. As in previous investigations of classical tearing mode saturation \citep{Loizu2020, Loizu_Bonfiglio_2023}, we make use of SPEC \citep{Hudson2012}, an equilibrium solver based on the variational principle of the Multi-Region relaxed MHD, featuring stepped pressure profiles and arbitrary magnetic topology. We work in slab geometry and employ a simple bootstrap current model $J_{bs} = C\:\bnabla p$ to study the bootstrap-driven tearing modes, scanning over the asymptotic matching parameter $\Delta'$ and bootstrap current strength. Saturated island widths produced by SPEC agree well with the predictions of an initial value resistive MHD code \citep{Huang_2016} while being orders of magnitude faster to calculate. Additionally, we observe good agreement with a simple analytical Modified Rutherford Equation, without requiring any fitting coefficients. The match is obtained for both linearly unstable classical tearing modes in the presence of bootstrap current, and neoclassical tearing modes, which are linearly stable but nonlinear-unstable due to the effects of the bootstrap current. 

\end{abstract}

\section{Introduction}

Tearing modes convert magnetic field energy into kinetic energy of the plasma through spontaneous magnetic reconnection \citep{Goldston2020}. Generally present in toroidally-confined magnetic fusion plasmas, these internal modes violate the frozen-in constraint of ideal MHD, modifying the flux surface topology and creating magnetic islands. In many cases, tearing modes do not lead to the complete collapse of the plasma, but rather grow into non-linearly saturated states \citep{Kong2020} that can strongly increase global transport and lower performance \citep{Gunter1999}. These instabilities often appear in the form of neoclassical tearing modes (NTM), the behavior of which is driven by the bootstrap current. The latter is generated by the collisional interaction of trapped and passing particles in a toroidal plasma and is typically proportional to pressure gradients \citep{Helander_Sigmar_2005}. As the transport of particles and heat along the magnetic field dominates that across the field, islands tend to locally flatten pressure \citep{Fitzpatrick1995}, which leads to the removal of bootstrap current inside the island, driving the tearing mode unstable. Effects of bootstrap current on tearing modes include an enhanced growth of linearly unstable tearing modes and, more importantly, a nonlinear destabilization of configurations that are linearly stable to tearing modes.

As a first approximation, tearing modes are studied using linear theory which involves tearing-like perturbations of an initial equilibrium. These are constructed by solving the linearized resistive MHD equations in a narrow layer around the resonant surface, where the safety factor $q$ is rational, and performing an asymptotic matching with a solution of the linearized ideal MHD equations outside the layer. If a mode can be found that lowers the MHD potential energy, $\mathcal{W}=\int dV\:[B^2/2\mu_0 + p/(\gamma-1)]$, then the equilibrium is deemed unstable to this mode. In the large aspect ratio limit, an important quantity is $\Delta'$ which comes from the discontinuity of the derivative of the perturbed flux across the resistive layer for a given perturbation. Neglecting the stabilizing effects of pressure \citep{ggj}, if $\Delta'>0$, then a particular tearing mode is unstable and will lower $\mathcal{W}$, and if $\Delta'<0$ the mode will increase $\mathcal{W}$ as it takes more energy to bend the fieldlines than what is released by reconnection. The linear approach provides a good start for uncovering the physics of tearing modes, but it only yields information on linear growth and stability. Therefore, it is of limited use for understanding nonlinear growth and saturation \citep{Goldston2020}, as is measured in experiments \citep{Kong2020}.

\begin{figure}
    \centering
    \includegraphics{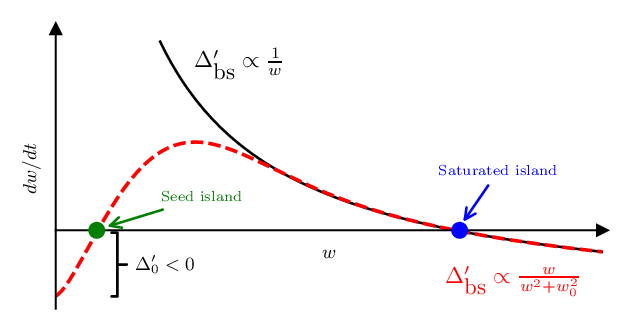}
    \caption{Sketch of the width growth rate as a function of width, as described by Equation \ref{eq1}. The black curve shows the behavior with the diverging $1/w$ term, while the red curve demonstrates the effect of the $w_0$ correction to $\Delta'_\textrm{bs}$.}
    \label{fig:mre}
\end{figure}

To study the nonlinear dynamics, \citet{Rutherford1973} started from the linear theory and extended it into a nonlinear regime, deriving an ODE for the evolution of the island width $w$, which is a good measure of the mode amplitude and also relates to the level of confinement degradation of the underlying equilibrium. \citet{Fitzpatrick1995} extended Rutherford's model by including the effects of the bootstrap current. An example of such ODE is:
\begin{equation}
\label{eq1}
  \frac{\mu_0}{\eta} \deriv{w}{t} = \Delta'_0 + \Delta'_\textrm{stab}(w) + \Delta'_\textrm{bs}(w)
\end{equation}
where $\Delta'_0$ is the value of $\Delta'$ for perturbations of an initial equilibrium with $w=0$, $\Delta'_\textrm{stab}(w) \leq 0$ includes stabilizing terms and ensures saturation of the island, and $\Delta'_\textrm{bs}(w)$ accounts for the effects of the bootstrap current. The bootstrap term is destabilizing and can be found analytically in the large aspect ratio limit, $\Delta'_\textrm{bs}(w)\sim1/w$ \citep{Fitzpatrick1995}. This term diverges for small islands and so is modified to $\Delta'_\textrm{bs}(w)\sim w^2/(w^2+w_0^2)$, thus introducing a regularizing scale, $w_0\sim({\chi_\perp}/{\chi_\parallel})^{1/4}$, which describes the effects of incomplete pressure flattening in an island. Here $\chi_\perp$ and $\chi_\parallel$ are the heat diffusion coefficients across and along the magnetic field, respectively. The typical evolution of an island for a linearly stable equilibrium ($\Delta'_0<0$) using either of the two bootstrap terms is shown in Figure \ref{fig:mre}. When $w_0$ isn't included, the mode is always unstable and will grow regardless of the initial island size. However, when $w_0$ is included and incomplete flattening is accounted for, then the mode will not grow unless it has been perturbed with a sufficient seed, i.e. if the initial island size is sufficiently large. Seeding dynamics are complicated \citep{Muraglia2021}, but it is important to note that, in the large island limit $w \gg w_0$ , the saturated island is essentially the same, regardless of whether the seeding mechanism is included or not. For the purposes of this paper, we study the $w \gg w_0$ regime, which is one of greater relevance to magnetic fusion experiments (in a typical tokamak $w_0\approx 1 \textrm{cm}$ while $w_\textrm{sat} \approx 4 \textrm{cm}$) \citep{OSauter_2002}. Non-linear equations such as Eq. \ref{eq1} are known as Modified Rutherford equations (MRE) and are often used in tokamak experiments \citep{Kong2020} as they are quick to evaluate and give good estimates for the saturated island widths. However, MRE often includes other $\Delta'(w)$ ad-hoc terms \citep{Kong_2022} which are device-specific and can feature many fitting coefficients.

A more accurate approach to studying nonlinear tearing modes is the time integration of resistive MHD equations, evolving a perturbed equilibrium into a steady state featuring a saturated island. Such a method gives a detailed description of tearing mode dynamics and can be expanded to add various non-MHD effects, but it is numerically very expensive \citep{Hoelzl_2021, Teng_2018}. The cost comes from the large separation between ideal and resistive timescales (measured by Lundquist number $S = \tau_\textrm{res} / \tau_\textrm{Alfven} \gg 1$), which requires long simulation times, and the separation between system size and tearing spatial scales, which requires fine meshes. In summary, given existing approaches, it is difficult to predict the saturation of tearing modes in a manner that is simultaneously fast, geometry-agnostic, and rooted in solid theory.

However, there may exist another approach based on energy principles that can remedy some of these stated drawbacks. The method was pioneered by \citet{CooperGraves2010}, who used the ideal MHD equilibrium solver VMEC \citep{vmec} to predict the saturated state of internal kink modes in a tokamak. The idea is that, in addition to the initial MHD equilibrium, one can think of the saturated state of an instability as yet another (lower energy) MHD equilibrium. Thus, if an equilibrium code exists where both the initial and saturated states are in the space of possible solutions, then by clever perturbations in the parameter space it may be possible to directly jump from the initial to the final state. In addition to ideal modes such as the internal kink, this approach proved valuable for finding nonlinear states of non-resonant modes \citep{Brunetti_2014}, where $q=m/n$ surface is not present. Here, nonlinear states from VMEC \citep{vmec} agreed well with a resistive MHD code when a mode was nonresonant, but led to a poor agreement otherwise. 

Loizu extended the method for resonant resistive instabilities. His work showed that such an approach can be used for the prediction of the saturated state of classical tearing modes, both in the slab \citep{Loizu2020} and cylindrical geometry \citep{Loizu_Bonfiglio_2023}. Tearing mode prediction required using an equilibrium code such as the Stepped Pressure Equilibrium Code (SPEC) \citep{Hudson2012}, which does not constrain the magnetic topology around the resistive layer. In this paper, we present an extension of the work done in slab \citep{Loizu2020} where the effects of bootstrap current are included and lead to the growth and saturation of neoclassical tearing modes. The paper is organized as follows. In section 2 we introduce the set of tearing-unstable initial equilibria, including the geometry and starting profiles. Section 3 describes how the saturation of tearing modes is predicted using resistive MHD initial value simulations. Finally, section 4 presents the study of nonlinear saturation using SPEC, along with a comparison of the variational approach to the resistive MHD and MRE. 

\section{Family of tearing unstable equilibria}

To demonstrate the new approach for the prediction of tearing mode saturation with bootstrap current, we work in slab geometry. This simplifies the analysis since the slab does not have curvature or a coordinate singularity. In addition, we can set up an equilibrium that is stable to ideal modes and that has only one resonant surface where a tearing mode can grow unstable. Furthermore, analytical linear and nonlinear theories are simpler in slab geometry. We use a cartesian coordinate system where $x$ is a radial-like coordinate, $y$ is the poloidal-like direction, $z$ is the toroidal-like direction, and the slab is doubly periodic in $y$ and $z$. The size of the slab is fixed to $2\pi$ along $x$ and $z$, while we consider different values $L$ of the system size along $y$, thus allowing us to scan different stability regimes, with effects similar to varying the safety factor $q$ in toroidal geometry. An initial equilibrium magnetic field $\vect{B}_0=B_{z0} \hat{z} + \hat{z}\times\bnabla\psi_0$ is constructed with a flux function $\psi_0(x)=1/\cosh^2{(x)} + \alpha x^2$, similar to that in \citet{Loureiro2005}. Associated to this flux function is the poloidal field, $B_{y0}(x)=\psi_0'(x)$, and the toroidal current, $J_{z0}(x)=\psi_0''(x)$, which we call the ohmic current. The initial pressure $p_0(x)$ is simply a linear function of $x$, with the scale being given by the condition on the average plasma beta $\beta=\langle 2\mu_0 p/B^2 \rangle \approx 3 \%$. The equilibrium, profiles of which are shown in Fig. 2, is ideally stable and has a single resonant surface at $x=0$, and $B_{z0}$ is an essentially constant strong guide field, such that the mode is approximately incompressible \citep{Goldston2020}, with minor variations to set up an initial force balance, $B_{z0}^2 + |\bnabla\psi_0|^2 + p_0 = \textrm{const}$. The field at the resonant surface is solely in the $z$ direction, which means that the mode, in order to have the resonance condition $\vect{k}\cdot\vect{B}=0$, is symmetric in the $z$ direction. The bootstrap current, which in a torus originates from the field curvature, is not inherently present in a slab equilibrium. We overcome this by employing a simple model for the bootstrap current $\vect{J}_\textrm{bs} = \hat{C}\: C_0 (\partial p/ \partial x) \:\hat{z}$,
where $C_0=J_{z0}(0)/(\partial p_0(0)/ \partial x)$ and $\hat{C}$ is a free constant which specifies the relative strength of the bootstrap and ohmic current in the initial equilibrium. This term has been successfully used in previous studies of slab NTMs \citep{Muraglia2021}. Thanks to the POEM theory \citep{Escande2004, Militello2004}, the stabilization term of MRE, Eq. \ref{eq1}, can be expressed analytically as $\Delta'_\textrm{stab}(w)=-(1/2.44d^2) w$, where $d$ is the length scale of the initial equilibrium current, $d=\sqrt{J_{z0}(0)/J_{z0}''(0)}\approx 0.35$. Using the POEM stabilization term and a simplified bootstrap contribution $\Delta'_{bs} = \hat{C}\: C_0\:p'_r\:b\:(1/w)$, where $p'_r$ is the pressure gradient across the resonant surface and $b$ is a fitting coefficient (which can be estimated analytically, see Appendix B), we arrive at the following expression
\begin{equation}
    w_{sat} = 1.22 d^2 \paran{\Delta'_0 + \sqrt{\Delta_0'^2 + 1.64a^{-2}\:\hat{C}\: C_0\:p'_r\:b}}
\end{equation}
This equation provides an estimate of the island width for tearing modes with small constant $J_\textrm{bs}$ and is valid only in the large island limit, $w \gg w_0$. 

\section{Resistive MHD}

The canonical model equations describing the nonlinear evolution of tearing modes are the resistive MHD equations:
\begin{gather}
    \deriv{\rho}{t} + \bnabla \vecdot \paran{\rho \vect{u}} = 0 \\
    \deriv{(\rho \vect{u})}{
t} + \bnabla \vecdot 
\paran{\rho \vect{u} \vect{u}} = - \bnabla \paran{p + \frac{1}{2} B^2} + \bnabla \vecdot \paran{ \vect{B} \vect{B}} + \bnabla \vecdot \paran{ \mu \rho \boldsymbol{\tensor{\epsilon}}} \\
    \deriv{p}{t} + \bnabla \vecdot \paran{p \vect{u}} =  (1-\gamma) \left(p \bnabla \vecdot \vect{u}  -\frac{\chi_\parallel}{2} \bnabla \vecdot \left( \rho \vect{b} \vect{b} \vecdot \bnabla (p/\rho) \right) \right) \\
    \deriv{\psi}{t} = (\vect{u} \times \vect{B})_z + \eta (J_{z0} - J_z) + \eta J_\textrm{bs}
\end{gather}
These equations are solved, in a dimensionless form, using the code HMHD \citep{Huang_2016}. Viscous forces are introduced in the momentum balance, Eq. 3.2, through the strain tensor $\boldsymbol{\tensor{\epsilon}}=(\bnabla \vect{u} + \bnabla^T \vect{u})/2$. In Ohm's law, Eq. 3.4, $\eta J_{z0}$ is the loop voltage which maintains a non-zero steady state, and $\eta J_\textrm{bs}$ acts like a secondary driving term, allowing for the inclusion of an effective bootstrap current, as used by \citet{Muraglia2021}. The variables are normalized as follows (real $\rightarrow$ normalized): length $x,y,z\rightarrow x,y,z /L_0$, density $\rho\rightarrow\rho/\rho_0$, magnetic field $\vect{B}\rightarrow \vect{B} / B_0$, current $\vect{J} \rightarrow \vect{J} / (B_0/L_0\mu_0)$, velocities $\vect{u}\rightarrow \vect{u}/ u_\textrm{A}$ (with Alfven speed $u_\textrm{A}=B_0/\sqrt{\mu_0 \rho_0}$), time $t\rightarrow t/ \tau_\textrm{A}$ (with Alfven time $\tau_\textrm{A}=L_0 / u_\textrm{A}$), and pressure $p \rightarrow p/(B_0^2 \mu_0)$. The parameters of the model are the adiabatic coefficient $\gamma$, which we fix to that of the ideal gas $\gamma=5/3$, parallel diffusion coefficient, $\chi_\parallel \approx 10$, used for numerical stability, dynamic viscosity $\mu$, and resistivity $\eta$. The latter two are specified in terms of their respective dimensionless numbers, the Lundquist number $S=L_0 u_\textrm{A}/\eta$, and the magnetic Prandtl number $Pr_{m} = \mu/\eta$. For the simulations shown below, the dimensionless numbers are $S \in [10^3, 10^4]$ and $Pr_m = [10^{-1}, 10^{1}]$. The code relies on the 5-point finite difference scheme for the spatial discretization, and the trapezoidal leapfrog method for the temporal discretization. The simulations are performed using dynamic time steps based on the CFL condition \citep{cfl} for Alfvén dynamics. The mesh comprises approximately 400x400 points, periodic boundary conditions are applied in the $y$ direction, and wall boundary conditions are applied in the $x$ direction.

\begin{figure}
    \centering
    \includegraphics{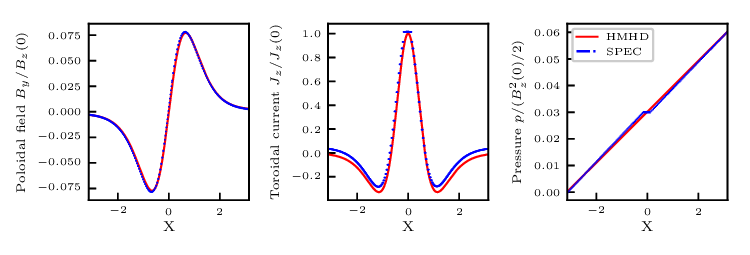}
    \vspace{-.4cm}
    \caption{Initial profiles ($\alpha=-0.002$) of the poloidal field, toroidal current, and pressure, as used in HMHD (red) and SPEC (blue). SPEC equilibrium consists of $N_{vol}=151$ volumes, the largest of which is the central volume, located around $x=0$.}
    \label{fig:initprof}
\end{figure}

\begin{figure}
    \centering
    \includegraphics{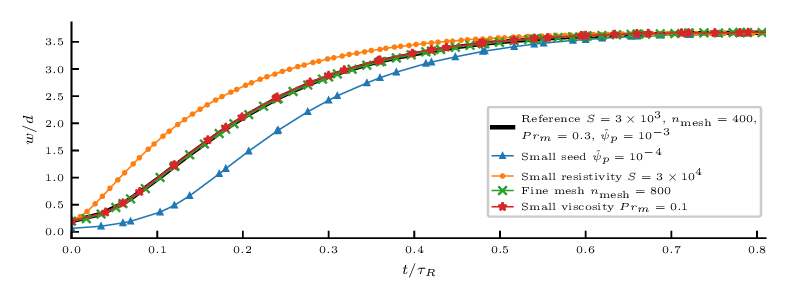}
    \vspace{-.4cm}
    \caption{Time evolution of the island width from a seed to a saturated island, as a function of the resistive time. Here the equilibrium has $L=10.7 d$ and $\Delta'_0\: d=0.97$. Different curves correspond to (1) a reference case (solid black), (2) a smaller seed (blue triangles), (3) smaller resistivity (orange circles), (4) finer mesh (green crosses), and (5) smaller viscosity (red stars).}
    \label{fig:hmhdwvstime}
\end{figure}

Simulating a tearing mode entails setting up the initial equilibrium, Figure \ref{fig:initprof}, according to the profiles derived from the initial flux function $\psi_0$ and pressure $p_0$. The initial density is constant and set to $\rho \in [2, 8]$ in normalized units. Note that the initial equilibrium does not include bootstrap current. The next step involves perturbing the initial flux with $\psi_p = \hat{\psi}_p \cos{(y/L) \cos{(x/2\pi)}} $, where $\hat{\psi}_p$ is the perturbation magnitude (the exact value does not affect saturation, see Fig. \ref{fig:hmhdwvstime}). The mode then grows spontaneously and reaches, on an approximately resistive time scale, $\tau_R={\mu_0 L^2}/{\eta}=S \tau_\textrm{Alfven}$, a state where the profiles and island size stabilize. The time trace of the width of a typical island is depicted in Figure \ref{fig:hmhdwvstime}. The final step is to fit a decaying exponential of the form $w(t)= c_0 + c_1 \exp(c_2 t)$ to the final stage of the island evolution. The best fit for the constant coefficient gives the saturated island width, $w_\textrm{sat}:=w(t\rightarrow \infty) = c_0$. Fig. 3 also shows that the value of $w_\textrm{sat}$ is independent of $\eta$, $\mu$, as well as the mesh resolution. The structure of the saturated island is shown in Figure 4a, where an $m=1$ mode can be seen to dominate. This supports an underlying assumption in MRE, which is that the nonlinear profiles have the same spatial dependence as the linear ones and are of single harmonicity. 

\begin{figure}
    \centering
    \includegraphics{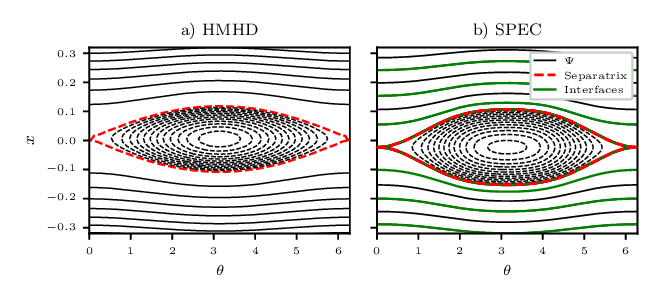}
    \vspace{-.4cm}
    \caption{Separatrices (red) and flux contours (black) of saturated islands in HMHD (left) and SPEC (right) for an equilibrium with $L=7.8 d$, which corresponds to $\Delta'_0\:d= -0.23$. The SPEC plot also depicts a selection of the ideal interfaces (green), the shape of which resembles the flux contours in HMHD.}
    \label{fig:islandplots}
\end{figure}

\begin{figure}
    \centering
    \includegraphics{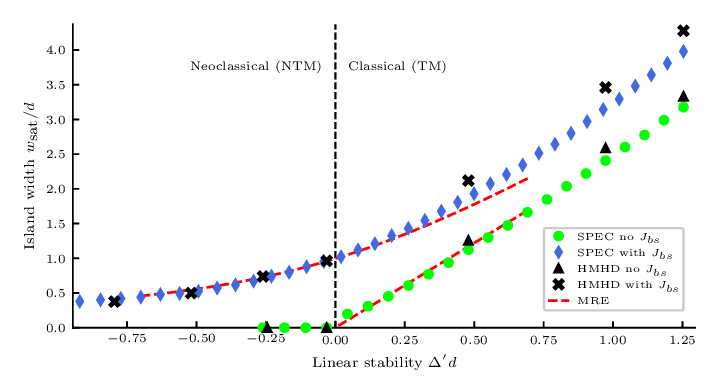}
    \vspace{-.4cm}
    \caption{Scan of island width as a function of the linear stability $\Delta'_0$. Classical modes are shown from SPEC (green circle), HMHD (black triangle). Bootstrap-enhanced classical modes and linearly stable neoclassical tearing modes are shown from SPEC (blue diamond) and HMHD (black cross). The analytical prediction from MRE is also displayed (red dash). The bootstrap coefficient used is $\hat{C}=0.048$.}
    \label{fig:masterplot}
\end{figure}

The islands shown in Figs. 3,4 are for a slab of a certain size $L$, to which corresponds a value of $\Delta'_0$. It is interesting to study the dependence of $w_\textrm{sat}$ on $\Delta'_0$. Such a study is shown in Figure \ref{fig:masterplot}, where the island width is depicted as a function of the linear stability parameter $\Delta'_0$. The black triangles reproduce the original study by \citet{Loizu2020}, showing that classical modes ($J_\textrm{bs}=0$) do not have a saturated state if linearly stable ($\Delta'_0<0$) and do have a saturated state, with $w_\textrm{sat}$ approximately linear in $\Delta'_0$, if linearly unstable ($\Delta'_0>0$). However, this is no longer true for tearing modes with bootstrap current, simulations of which are shown as black crosses (for $\hat{C}=0.048$). In this case, linearly stable modes ($\Delta'_0<0$) can in fact grow into saturated islands. These are usually referred to as neoclassical tearing modes (NTM) and are commonly observed in tokamaks \citep{Kong2020}. As depicted in Figure \ref{fig:mre}, NTMs need to be initiated with a sufficiently large seed, otherwise islands collapse back into the initial equilibrium. The precise size of this seed, however, does not impact the saturated island (see Fig. 3). For linearly unstable equilibria ($\Delta'_0>0$), the bootstrap current again acts in a destabilizing manner, resulting in an enhanced island width as compared to the cases with $J_\textrm{bs}=0$. 

\section{Variational approach}

\begin{figure}
    \centering
    \includegraphics{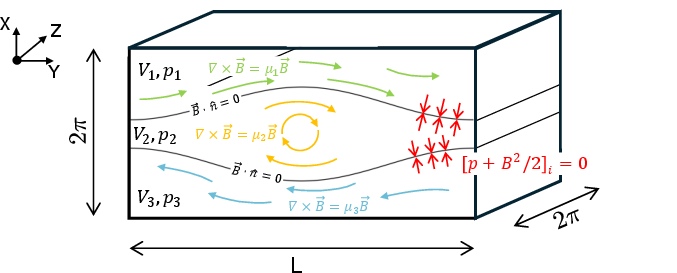}
    \caption{The setup for SPEC in slab geometry for a case of $N_\textrm{vol}=3$. SPEC solves for Beltrami fields within each volume and seeks force balance across ideal interfaces by perturbing the shapes of these interfaces.}
    \label{fig:specslab}
\end{figure}

For directly predicting the saturation of tearing modes, we use SPEC \citep{Hudson2012}, a code that finds Multi Region Relaxed MHD (MRxMHD) equilibrium states using a variational principle. 
\citet{Taylor1974} hypothesized that resistive plasmas often evolve in a way that conserves global magnetic helicity $\mathcal{H}=\int_{V_p} dV \vect{A}\vecdot\vect{B}$ while minimizing plasma energy $\mathcal{W}$. Here $\vect{A}$ is the vector potential and $V_p$ is the plasma volume. \citet{HOLE_HUDSON_DEWAR_2006} extended this idea to build the MRxMHD model that subdivides a plasma into $N_\textrm{vol}$ constant-pressure Taylor-relaxed volumes $\mathcal{V}_l$ which are separated by ideal interfaces $\mathcal{I}_l$. Thus SPEC solves MRxMHD by finding the extrema of the energy functional
\begin{equation*}
    \mathcal{F} = \mathcal{W} + \mu (\mathcal{H}-\mathcal{H}^0) = \sum_l \int_{\mathcal{V}_l} dV\: \paran{\frac{B^2}{2\mu_0}+\frac{p_l}{\gamma-1}} + \mu_l \paran{ \int_{\mathcal{V}_l}  dV\:\vect{A}\vecdot\vect{B} - \mathcal{H}^0_{l}}
\end{equation*}
where $\mathcal{H}^0_l$ are the target helicities and $\mu_l$ are the Lagrange multipliers. In slab geometry, the interfaces $\mathcal{I}_l$ are represented as
\begin{equation*}
    \vect{x}_{l}(\theta, \zeta) = \sum_m x_{l, m} \cos{(m\theta)}\:\hat{x} + L \frac{\theta}{2\pi}\:\hat{y} + \zeta\:\hat{z}
\end{equation*}
where $x_{l, m}$ are the coefficients of a Fourier expansion that is numerically truncated at $m=m_\textrm{pol}$. The field within each subvolume is found through minimization with respect to the vector potential $\vect{A}$, and force balance across the interfaces is achieved by varying the interface harmonics $x_{l, m}$. SPEC finds the states that extremize $\mathcal{F}$, which satisfy \citep{Hudson2012}
\begin{gather}
    \bnabla\times\vect{B} = \mu_l \vect{B} \\
    \boldsymbol{F}_l = [p + B^2/2\mu_0]_l = 0
\end{gather}
where the first equation describes linear force-free fields inside each $\mathcal{V}_l$, and the second equation represents the force balance across the two sides of each $\mathcal{I}_l$. The position and shape of each interface is in fact determined by Eq. 4.2, which is the local equivalent of the MHD force balance equation $\vect{J}\times\vect{B} = \bnabla p$. The standard approach for finding minimal MRxMHD states uses a Newton method for finding the zero of $\boldsymbol{F}_l$ and the alternative is a gradient descent algorithm for minimizing the functional $\mathcal{F}$. We found that the Newton method alone is not robust enough for equilibria with islands. Instead, a more robust approach involves first minimizing the energy functional using a descent method, which is slower but more stable, and then using the Newton method to make fine adjustments to interface harmonics, resulting in an equilibrium with good force balance. The setup for solving MRxMHD in slab geometry with SPEC for a case with $N_\textrm{vol}=3$ is shown in Figure \ref{fig:specslab}.

Inputs to SPEC include four physical constraints per volume that are fixed during functional minimization. These are the pressure $p_l$, the integrated toroidal and poloidal fluxes, $\Delta\Psi^t_l$ and $\Delta\Psi^p_l$, and the helicities $\mathcal{H}_l$ inside each volume. Alternatively, SPEC can directly solve Eqs. 4.1 and 4.2 provided a different set of four constraints, ($p_l$, $\Delta\Psi^t_l$, $\Delta\Psi^p_l$, $\mu_l$) or ($p_l$, $\Delta\Psi^t_l$, $I^\textrm{vol}_l$, $I^\textrm{surf}_l$), where $I^\textrm{vol}_l$ is the integrated toroidal current in each volume and $I^\textrm{surf}_l$ is the sheet current on each ideal interface \citep{Baillod_2021}. 

While the linear growth of tearing modes occurs on a time scale that is intermediate between Alfvenic, $\tau_\textrm{Alfven}$, and resistive, $\tau_R$ (for slab $\tau=\tau_\textrm{Alfven}^{2/5}\tau_R^{3/5}$ \citep{Goldston2020}), the nonlinear saturation occurs on an approximately resistive time scale, as is demonstrated in Figure \ref{fig:hmhdwvstime}. Therefore, Taylor's hypothesis of global helicity conservation, which works well for faster relaxation events \citep{Berger1999}, is not valid for slow tearing modes. Instead, the POEM theory \citep{Militello2004} reveals that a good invariant is the flux surface average of the toroidal current $J_z(\psi) = \langle J_{z0} \rangle_\psi$, with $\psi$ being the nonlinear flux function. See \citet{Loizu2020} for a numerical demonstration of this invariant and its comparison with helicity. To construct equilibria where an integrated version of $J_z(\psi) = \langle J_{z0} \rangle_\psi$ is fixed, we run SPEC with the constraints ($p_l$, $\Delta\Psi^t_l$, $I^\textrm{vol}_l$, $I^\textrm{surf}_l$).

As with HMHD, finding tearing modes in SPEC involves first setting up the equilibrium. The initial domain is covered along the $x$ direction with ideal interfaces, with a number $N_\textrm{vol}$ that is sufficient to reproduce the equilibrium profiles. Special care needs to be given to the placement of the two interfaces bounding the resonant surface $x=0$. If these are placed too closely to the resonant surface then they might limit the growth of the island. Conversely, if the spacing between them is too large then the analytical profiles are not well represented around $x=0$, and can be artificially stabilized. The arbitrariness in interface placement is removed by placing the two interfaces symmetrically around $x=0$ with a spacing of $\Psi_w$ (defined as the normalized toroidal flux enclosed by these two interfaces) which maximizes the width of the resulting saturated island. For more details on interface placement, see \citet{Loizu2020}. From the interface positions and the analytical equilibrium profiles, the SPEC input profiles including $\Delta\Psi^t_l$, $I_l^\textrm{vol}$, and $p_l$ are calculated. Since the pressure in SPEC is stepped, the bootstrap model $J_\textrm{bs}=\hat{C}\:C_0\: \partial p/ \partial x$ is implemented as a set of sheet currents on the interfaces $I_l^\textrm{surf} = \hat{C}\:C_0\: \Delta p_{i}$. The integrated bootstrap current in SPEC is the same as in HMHD, and, as we will show, equivalent saturated islands can be obtained despite the discreteness of the current sheets. Note that the bootstrap current in SPEC, unlike in HMHD, exists already in the initial equilibrium, which explains why the initial toroidal volume currents, shown in Fig. 2, are not exactly the same between the two codes. SPEC is run with these inputs to find the initial MRxMHD state, profiles of which are shown in Fig. 2. 

The next step is to evaluate linear stability and determine the unstable eigenmode with which the initial equilibrium can be perturbed, before a secondary equilibrium with an island is sought. This is accomplished by analyzing the Hessian matrix \citep{Kumar_2022} defined as
\begin{equation*}
    \tensor{H}_{ij} = \deriv{\mathcal{F}}{\vect{X}_i\: \partial \vect{X}_j}
\end{equation*}
where $i, j$ index the packed degrees of freedom of the interfaces $\vect{X} = \{ x_{l,m} \:|\: l \!=\! 1, ...,  N_\textrm{vol}; \: m \!=\! 0, ..., m_\textrm{pol} \}$. Eigendecomposition of $\tensor{H}$ gives information on linear stability, where a negative eigenvalue implies instability and the associated eigenmode describes the shape of the mode in the space of interface Fourier coefficients. Since the initial equilibrium is ideally stable, depending on $L$, there is either zero or one negative eigenvalue, the eigenmode of which has the shape of an odd-parity tearing-like mode. Linear stability in SPEC has been verified for classical tearing modes \citep{Loizu2020}, while the stability of modes with bootstrap current is more subtle. In fact, as discussed in Sec. 1 and illustrated in Fig. 1, a system for which $w_0\rightarrow0$ (which corresponds to perfect pressure flattening in the island) is unconditionally unstable. Since the pressure profile in SPEC is, by definition, flat in the relaxation region, we expect that tearing modes will always be unstable when bootstrap effects are included. This is in fact what we observe when looking at the Hessian (see Appendix A). Nevertheless, as discussed in Sec. 1 and seen in Fig. 1, the saturation of an island is essentially independent of $w_0$ in the limit $w \gg w_0$ studied here. After finding the eigendecomposition of the Hessian, the next step involves perturbing the interfaces of the SPEC equilibrium, creating the seed island. The perturbation is constructed from the unstable eigenmode which is scaled to give the largest possible seed that does not overlap the ideal interfaces. Finally, SPEC is run for the second time, iterating on the interface harmonics to reach a secondary MRxMHD equilibrium with a lower energy and a topology of an island. 

The results of running SPEC for a range of equilibria can be seen in Figure \ref{fig:masterplot}. The green circles correspond to classical tearing modes without bootstrap current, as reported in \citet{Loizu2020}. In this case, SPEC recovers the island width obtained using HMHD as well as the prediction by the POEM theory \citep{Militello2004}. As is shown, SPEC correctly predicts the linear stability threshold such that stable equilibria for which $\Delta'_0<0$ are also nonlinearly stable. The blue diamonds in Fig. 5 show the results from running SPEC with the inclusion of bootstrap current ($\hat{C}=0.048$). SPEC again manages to predict islands both in the case of classical modes (enhanced with bootstrap) and neoclassical tearing modes (which are only destabilized by the bootstrap current effect). A slight difference exists in the width between SPEC and HMHD for the larger islands, which we believe may be due to a violation of the current invariant with the increasing island size. Also shown in the figure is a prediction of an MRE, Eq. 2.1, which includes the bootstrap contribution. The results related to the MRE rely on a constant that we fit numerically, $b=4.25$, as is standard for these models in general \citep{Kong2020}, instead of using the less-accurate, analytical value of $b=8$, as derived in Appendix \ref{appB}. The fit is obtained by matching the island width for the equilibrium with $\Delta'_0=0$.

\begin{figure}
    \centering
    \includegraphics{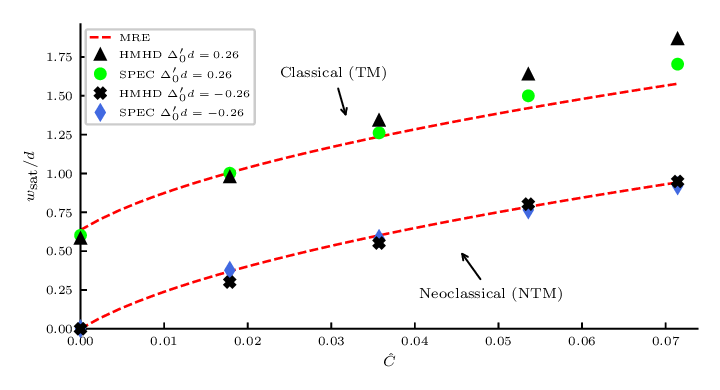}
    \vspace{-.4cm}
    \caption{Scan of island width as a function of the bootstrap current strength for SPEC (green and blue) and HMHD (black), in the case of a classical tearing mode ($\Delta'_0\: d =0.26$) and an NTM ($\Delta'_0\: d=-0.26$).}
    \label{fig:csan}
\end{figure}

The analysis in Figure \ref{fig:masterplot} was done using a fixed bootstrap strength, $\hat{C}=0.048$. To verify that the prediction from SPEC is independent of the exact value of $\hat{C}$, we chose two equilibria, one linearly stable, $\Delta'_0\:d=-0.26$, and one linearly unstable, $\Delta'_0\:d=0.26$, and we vary the constant $\hat{C}$ from $0$ to $0.3$. The resulting island widths are shown in Figure \ref{fig:csan}, where it can be seen that the value of $\hat{C}$ does not significantly affect the agreement between SPEC, HMHD, and the analytical MRE. In addition to the island width, one can look at the structure of the island in SPEC and compare it with the island generated in HMHD. As shown in Figure 4, the island from SPEC matches that of HMHD closely. Its shape is dominantly given by the $m=1$ harmonic, but it seems to have a slightly larger content of $m>1$ harmonics. It is interesting to note that the two interfaces around the island, shown in green in Fig. 4b, are effectively coincident on the island separatrix, such that the entire volume between them is filled by the island. This is a consequence of the choice of $\Psi_w$ that maximizes the island width.

\begin{figure}
    \centering
    \includegraphics{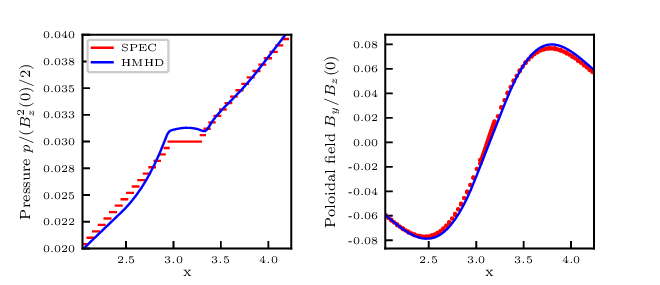}
    \vspace{-.4cm}
    \caption{Profiles of a) pressure and b) poloidal field for a saturated tearing mode from SPEC and HMHD. Flattened pressure in HMHD matches well with SPEC where the pressure is, ipso facto, flat inside the island. The poloidal field is mostly unchanged from the initial equilibrium.}
    \label{fig:satprof}
\end{figure}

The saturated profiles in SPEC also agree well with the profiles from HMHD, as can be seen in Figure \ref{fig:satprof}. The pressure in HMHD flattens across the island, which in this case happens primarily due to the generated parallel plasma flow, and not the parallel diffusion. In SPEC, the flattening is a consequence of the MRxMHD model. The radial profile of $B_y$ over the entire domain is not much affected by the tearing mode, meaning that the initial agreement between SPEC and HMHD is sustained in the saturated state. 

Two important parameters in SPEC are $m_\textrm{pol}$ and $N_\textrm{vol}$. The first one determines the number of poloidal harmonics used for representing the interfaces and force imbalances, while the second one is the number of relaxed volumes in SPEC and it determines how well the outer ideal solution is represented. In Figure 9, the saturation width of a classical tearing mode obtained with SPEC is shown as a function of both $m_\textrm{pol}$ and $N_\textrm{vol}$. It can be seen that the saturated island width is converged, in the sense that increasing the two parameters beyond $m_\textrm{pol}=3$ and $N_\textrm{vol}=151$, which are used for the simulations presented in this paper, leads to less than $1\%$ difference in the calculated widths.

\begin{figure}
    \centering
    \includegraphics{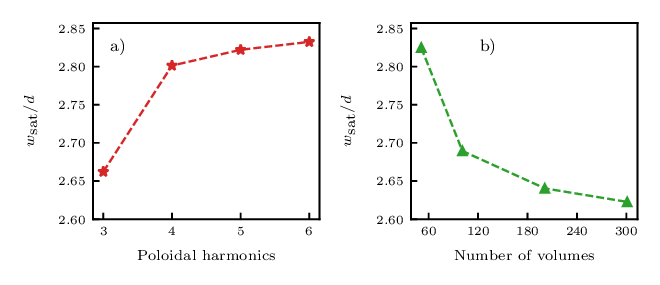}
    \vspace{-.4cm}
    \caption{Convergence of the width of a classical tearing mode with the bootstrap current in SPEC with respect to the number of a) poloidal harmonics and b) relaxed volumes. Reaching an equilibrium with a saturated island requires at least 3 harmonics.}
    \label{fig:specconv}
\end{figure}

\section{Discussion}

We have demonstrated that an "equilibrium approach", based on the solver SPEC \citep{Hudson2012}, can be used to directly predict the saturated state of tearing modes that evolve under the influence of the neoclassical bootstrap current. Good agreement with the resistive MHD equations, solved using the code HMHD \citep{Huang_2016}, was found for linearly unstable classical tearing modes, where the bootstrap current enhances the width of the saturated island. More importantly, we have shown that the approach can be applied to neoclassical tearing modes, which are linearly stable and for which bootstrap current is the primary driving mechanism. In addition to demonstrating the agreement in the width of saturated islands, a good match was also shown for the saturated island shapes and profiles of the pressure and the magnetic field. The independence of the SPEC results with respect to resolution parameters was established, as shown by the scans of the island width as a function of the number of poloidal harmonics and relaxed volumes. Having demonstrated that SPEC can correctly reproduce the saturated state of tearing modes, it is important to stress the relative benefits of this approach. First and foremost, the new method is significantly faster than resistive MHD simulations. For the scan of tearing modes with the bootstrap current in Figure \ref{fig:masterplot}, it took approximately 300 CPU hours to simulate with HMHD the saturation of just a single tearing mode. Conversely, it took a mere 2 CPU hours to simulate the entire dataset for SPEC, which includes 30 different saturated islands. An additional benefit is that the SPEC-based approach does not involve fitting coefficients. Such is the case with the analytical MRE model which, despite being fast, has several coefficients that need to be fit to a specific device ($\sim10$ coefficients for the TCV tokamak) \citep{Kong2020}. The last notable benefit of the approach is that it is general, both in terms of geometry and particular instability. While initial work \citep{Loizu2020} and its extension presented here were done in a slab, this approach could be applied to other configurations. The method has already been applied to classical tearing modes in a cylindrical tokamak \citep{Loizu_Bonfiglio_2023}, and work has started in toroidal geometry. We are interested in predicting 2-1 NTMs observed in the TCV tokamak \citep{Kong2020} and comparison with experimental measurements and MRE models. More broadly, this work promotes the development of a fast and simple tool for the general prediction of saturated instabilities in arbitrary geometries. In particular, this method may eventually allow for quick prediction of nonlinear saturation in stellarators, which have been observed \citep{deAguilera_2015, Watanabe_2005, Geiger2004} to operate stably in regimes deemed unstable by linear theory. 

\section{Acknowledgments}

This work was supported in part by the Swiss National Science Foundation. And by a grant from the Simons Foundation (1013657, JL). This work has been carried out within the framework of the EUROfusion Consortium, via the Euratom Research and Training Programme (Grant Agreement No 101052200 — EUROfusion) and funded by the Swiss State Secretariat for Education, Research and Innovation (SERI). Views and opinions expressed are however those of the author(s) only and do not necessarily reflect those of the European Union, the European Commission, or SERI. Neither the European Union nor the European Commission nor SERI can be held responsible for them. We thank Antoine Baillod for his help with the codes and insightful discussions.

\appendix

\section{}\label{appA}

\begin{figure}
    \centering
    \includegraphics{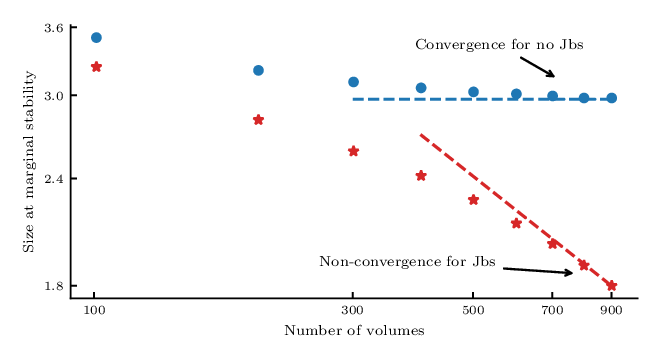}
    \vspace{-.4cm}
    \caption{Size of a marginally stable slab in SPEC as a function of the number of volumes. The top curve (blue) is for the case with $J_\textrm{bs}=0$, which converges to a well-defined value. The bottom curve (red) has $J_\textrm{bs}\neq0$ and it does not have a limit.}
    \label{fig:specstab}
\end{figure}

Linear stability in SPEC has been verified against numerical calculations of $\Delta'_0$ for classical tearing modes \citep{Loizu2020}. The stability for equilibria with bootstrap current is more involved. In Figure \ref{fig:specstab}, we measure the size of a marginally stable slab as a function of the number of relaxed volumes. Each set of points in the plot represents marginal stability, such that points above are unstable and below stable. The top curve in blue represents equilibria without bootstrap current, and it can be seen that it has good convergence as $N_\textrm{vol}$ is increased, in this case to a marginal size of $\sim 3.0$. However, the same is not true for the equilibria with bootstrap current shown in red, for which the limit $N_\textrm{vol}\rightarrow\infty$ does not seem to exist. This leads us to believe that such equilibria are linearly unconditionally unstable in SPEC, where stability in any particular equilibria is due to limited resolution. Such a conclusion is supported by the bootstrap term from the MRE (see Figure \ref{fig:mre}). In SPEC, the ratio of parallel to perpendicular heat conductivity is infinite, which means that the term that accounts for heat transport in the island becomes $w_0=0$. This leads to the bootstrap contribution becoming $\Delta'_\textrm{bs}\sim 1/w$, which also leads to unconditional linear stability in the MRE model.

\section{}\label{appB}

Here we derive the MRE for the width of tearing modes in slab with bootstrap current. Neglecting flows which are assumed to be weak, we start from Ohm's law
\begin{equation}
    \deriv{\psi}{t} = \eta (J_{z0} - J_z) + \eta J_\textrm{bs}
\end{equation}
Expressing the flux function as a sum of the equilibrium and perturbed part, $\psi(x, y, t) = \psi_0(x) + \psi_1(x, y, t)$, and assuming gradients in $x$ dominate over those in $y$, $\partial/\partial x \gg \partial/\partial y$,
\begin{equation}
    \deriv{\psi_1}{t} = \eta \deriv{^2\psi_1}{x^2} + \eta J_\textrm{bs}
\end{equation}
Assuming all quantities are of a single harmonicity in the poloidal $y$ direction, $k=2\pi/L$, the functions are expressed as $f(x,y,t) \sim \hat{f}(x,t)\cos(k y)$. Next, we integrate both sides from $-w/2$ to $w/2$, where $w$ is the island width at time $t$
\begin{equation}
    \int_{-w/2}^{w/2} dx\: \deriv{\psi_1}{t} = \int_{-w/2}^{w/2} dx\: \eta \deriv{^2\psi_1}{x^2} + \int_{-w/2}^{w/2} dx\: \eta J_\textrm{bs}
\end{equation}
For the term on the left hand side, we apply the constant-$\psi$ approximation to lift $\psi$ outside the integral, and for the bootstrap term, we assume that the pressure gradient, and hence $J_\textrm{bs}$, is slowly varying across the island. We obtain
\begin{equation}
    \deriv{\psi_1}{t} w = \eta \deriv{\psi_1}{x}\Bigr|_{-w/2}^{w/2} + \eta J_\textrm{bs} w
\end{equation}
To proceed further, we need an additional equation that relates the perturbed flux and instantaneous island width. Following an approach similar to \citet{Goldston2020}, the total flux is again composed of equilibrium and perturbed contributions. Expanding around the resonant surface, we use first-order expansion for the field $B_{y0}(x)=B'_{y0}x$, which gives $\psi_0=-1/2\:B'_{y0} x^2$, since we know that $B_{y0}(x)=-{\partial\psi_0}/{\partial x}$. For the perturbed flux, we use the zeroth-order expansion in $x$ and the fact that the island is of a single harmonicity $\psi_1(x,y,t)=-\hat{\psi}\cos(k y)$. 

Evaluating the flux function at the X-point, we get $\psi(x=0,y=0)=-\hat{\psi}$, and the following expression traces the field line ($x_l$, $y_l$) corresponding to that flux
\begin{equation}
    \hat{\psi} (1-\cos(k y_l)) = 1/2\: B'_{y0} x^2_l
\end{equation}
The island width is then obtained from the maximum excursion of the field line in $x$
\begin{equation}
    w(t) = 2 \max(x_l) = \left[ \frac{16\hat{\psi}}{B'_{y0}} \right]^{1/2}
\end{equation}
Thus we obtain the relation between the reconnected flux and the island width
\begin{equation}
    \psi = b w^2 \:\:\textrm{         with } b=\frac{B'_{y0}}{16}
\end{equation}
Finally, this expression is inserted into the relation Eq. B4, leading to the MRE model
\begin{equation}
    \frac{2}{\eta} \deriv{w}{t} = \Delta' + \frac{1}{b} J_\textrm{bs} \frac{1}{w}
\end{equation}
\begin{equation}
    \textrm{with } \Delta' = \frac{1}{\psi_1} \deriv{\psi_1}{x}\Bigr|_{-w/2}^{w/2}
\end{equation}
\bibliographystyle{jpp}

\bibliography{main}

\begin{thebibliography}{31}
\expandafter\ifx\csname natexlab\endcsname\relax\def\natexlab#1{#1}\fi
\def\au#1{#1} \def\ed#1{#1} \def\yr#1{#1}\def\at#1{#1}\def\jt#1{\textit{#1}} \def\bt#1{#1}\def\bvol#1{\textbf{#1}} \def\vol#1{#1} \def\pg#1{#1} \def\publ#1{#1}\def\arxiv#1{#1}\def\org#1{#1}\def\st#1{\textit{#1}}

\bibitem[de~Aguilera {\em et~al.\/}(2015)de~Aguilera, Castejón, Ascasíbar, Blanco, la~Cal, Hidalgo, Liu, López-Fraguas, Medina, Ochando, Pastor, Ángeles Pedrosa, Milligen, Velasco \& the TJ-II~Team]{deAguilera_2015}
{\sc \au{de~Aguilera, Adriana~M.}, \au{Castejón, Francisco}, \au{Ascasíbar, Enrique}, \au{Blanco, Emilio}, \au{la~Cal, Eduardo~De}, \au{Hidalgo, Carlos}, \au{Liu, Bing}, \au{López-Fraguas, Antonio}, \au{Medina, Francisco}, \au{Ochando, María~Antonia}, \au{Pastor, Ignacio}, \au{Ángeles Pedrosa, María}, \au{Milligen, Boudewijn~Van}, \au{Velasco, José~Luis} \& \au{the TJ-II~Team}} \at{ \yr{2015} }  \bvol{55}~(11),  \pg{113014}.

\bibitem[Baillod {\em et~al.\/}(2021)Baillod, Loizu, Qu, Kumar \& Graves]{Baillod_2021}
{\sc \au{Baillod, A.}, \au{Loizu, J.}, \au{Qu, Z.S.}, \au{Kumar, A.} \& \au{Graves, J.P.}} \yr{2021}  \at{Computation of multi-region, relaxed magnetohydrodynamic equilibria with prescribed toroidal current profile}.  \jt{Journal of Plasma Physics}  \bvol{87}~(4),  \pg{905870403}.

\bibitem[Berger(1999)]{Berger1999}
{\sc \au{Berger, Mitchell~A}} \yr{1999}  \at{Introduction to magnetic helicity}.  \jt{Plasma Physics and Controlled Fusion}  \bvol{41}~(12B),  \pg{B167}.

\bibitem[Brunetti {\em et~al.\/}(2014)Brunetti, Graves, Cooper \& Terranova]{Brunetti_2014}
{\sc \au{Brunetti, D.}, \au{Graves, J.P.}, \au{Cooper, W.A.} \& \au{Terranova, D.}} \yr{2014}  \at{Ideal saturated mhd helical structures in axisymmetric hybrid plasmas}.  \jt{Nuclear Fusion}  \bvol{54}~(6),  \pg{064017}.

\bibitem[Cooper {\em et~al.\/}(2010)Cooper, Graves, Pochelon, Sauter \& Villard]{CooperGraves2010}
{\sc \au{Cooper, W.~A.}, \au{Graves, J.~P.}, \au{Pochelon, A.}, \au{Sauter, O.} \& \au{Villard, L.}} \yr{2010}  \at{Tokamak magnetohydrodynamic equilibrium states with axisymmetric boundary and a 3d helical core}.  \jt{Phys. Rev. Lett.}  \bvol{105},  \pg{035003}.

\bibitem[Courant {\em et~al.\/}(1967)Courant, Friedrichs \& Lewy]{cfl}
{\sc \au{Courant, R.}, \au{Friedrichs, K.} \& \au{Lewy, H.}} \yr{1967}  \at{On the partial difference equations of mathematical physics}  \bvol{11}~(2),  \pg{215--234}.

\bibitem[Escande \& Ottaviani(2004)]{Escande2004}
{\sc \au{Escande, D.F.} \& \au{Ottaviani, M.}} \yr{2004}  \at{Simple and rigorous solution for the nonlinear tearing mode}.  \jt{Physics Letters A}  \bvol{323}~(3),  \pg{278--284}.

\bibitem[Fitzpatrick(1995)]{Fitzpatrick1995}
{\sc \au{Fitzpatrick, Richard}} \yr{1995}  \at{Helical temperature perturbations associated with tearing modes in tokamak plasmas}.  \jt{Physics of Plasmas}  \bvol{2},  \pg{825--838}.

\bibitem[Geiger J.~E.(2004)]{Geiger2004}
{\sc \au{Geiger J.~E., Weller, A. Zarnstorff M. C. Nührenberg C. Werner A. H. F. Kolesnichenko Y. I.~{W7-AS Team and Neutral Beam Injection Group}}} \yr{2004}  \at{Equilibrium and stability of high-b plasmas in wendelstein 7-as}.  \jt{Fusion Science and Technology}  \bvol{46}~(1),  \pg{13--23}.

\bibitem[Glasser {\em et~al.\/}(1976)Glasser, Greene \& Johnson]{ggj}
{\sc \au{Glasser, A.~H.}, \au{Greene, J.~M.} \& \au{Johnson, J.~L.}} \yr{1976}  \at{{Resistive instabilities in a tokamak}}.  \jt{The Physics of Fluids}  \bvol{19}~(4),  \pg{567--574}.

\bibitem[Goldston(2020)]{Goldston2020}
{\sc \au{Goldston, R.J}} \yr{2020} {\em Introduction to Plasma Physics\/}.  \publ{CRC Press}.

\bibitem[Günter {\em et~al.\/}(1999)Günter, Gude, Maraschek, Yu \& the ASDEX Upgrade~Team]{Gunter1999}
{\sc \au{Günter, S}, \au{Gude, A}, \au{Maraschek, M}, \au{Yu, Q} \& \au{the ASDEX Upgrade~Team}} \yr{1999}  \at{Influence of neoclassical tearing modes on energy confinement}.  \jt{Plasma Physics and Controlled Fusion}  \bvol{41}~(6),  \pg{767}.

\bibitem[Helander \& Sigmar(2005)]{Helander_Sigmar_2005}
{\sc \au{Helander, Per} \& \au{Sigmar, D.~J.}} \yr{2005} {\em Collisional transport in magnetized plasmas\/}.  \publ{Cambridge University Press}.

\bibitem[Hirshman \& Whitson(1983)]{vmec}
{\sc \au{Hirshman, S.~P.} \& \au{Whitson, J.~C.}} \yr{1983}  \at{{Steepest‐descent moment method for three‐dimensional magnetohydrodynamic equilibria}}.  \jt{The Physics of Fluids}  \bvol{26}~(12),  \pg{3553--3568}.

\bibitem[Hoelzl {\em et~al.\/}(2021)Hoelzl, Huijsmans, Pamela, Bécoulet, Nardon, Artola, Nkonga, Atanasiu, Bandaru, Bhole, Bonfiglio, Cathey, Czarny, Dvornova, Fehér, Fil, Franck, Futatani, Gruca, Guillard, Haverkort, Holod, Hu, Kim, Korving, Kos, Krebs, Kripner, Latu, Liu, Merkel, Meshcheriakov, Mitterauer, Mochalskyy, Morales, Nies, Nikulsin, Orain, Pratt, Ramasamy, Ramet, Reux, Särkimäki, Schwarz, Verma, Smith, Sommariva, Strumberger, van Vugt, Verbeek, Westerhof, Wieschollek \& Zielinski]{Hoelzl_2021}
{\sc \au{Hoelzl, M.}, \au{Huijsmans, G.T.A.}, \au{Pamela, S.J.P.}, \au{Bécoulet, M.}, \au{Nardon, E.}, \au{Artola, F.J.}, \au{Nkonga, B.}, \au{Atanasiu, C.V.}, \au{Bandaru, V.}, \au{Bhole, A.}, \au{Bonfiglio, D.}, \au{Cathey, A.}, \au{Czarny, O.}, \au{Dvornova, A.}, \au{Fehér, T.}, \au{Fil, A.}, \au{Franck, E.}, \au{Futatani, S.}, \au{Gruca, M.}, \au{Guillard, H.}, \au{Haverkort, J.W.}, \au{Holod, I.}, \au{Hu, D.}, \au{Kim, S.K.}, \au{Korving, S.Q.}, \au{Kos, L.}, \au{Krebs, I.}, \au{Kripner, L.}, \au{Latu, G.}, \au{Liu, F.}, \au{Merkel, P.}, \au{Meshcheriakov, D.}, \au{Mitterauer, V.}, \au{Mochalskyy, S.}, \au{Morales, J.A.}, \au{Nies, R.}, \au{Nikulsin, N.}, \au{Orain, F.}, \au{Pratt, J.}, \au{Ramasamy, R.}, \au{Ramet, P.}, \au{Reux, C.}, \au{Särkimäki, K.}, \au{Schwarz, N.}, \au{Verma, P.~Singh}, \au{Smith, S.F.}, \au{Sommariva, C.}, \au{Strumberger, E.}, \au{van Vugt, D.C.}, \au{Verbeek, M.}, \au{Westerhof, E.}, \au{Wieschollek, F.} \& \au{Zielinski, J.}} \yr{2021}  \at{The jorek non-linear extended
  mhd code and applications to large-scale instabilities and their control in magnetically confined fusion plasmas}.  \jt{Nuclear Fusion}  \bvol{61}~(6),  \pg{065001}.

\bibitem[Hole {\em et~al.\/}(2006)Hole, Hudson \& Dewar]{HOLE_HUDSON_DEWAR_2006}
{\sc \au{Hole, M.~J.}, \au{Hudson, S.~R.} \& \au{Dewar, R.~L.}} \yr{2006}  \at{Stepped pressure profile equilibria in cylindrical plasmas via partial taylor relaxation}.  \jt{Journal of Plasma Physics}  \bvol{72}~(6),  \pg{1167–1171}.

\bibitem[Huang \& Bhattacharjee(2016)]{Huang_2016}
{\sc \au{Huang, Yi-Min} \& \au{Bhattacharjee, A.}} \yr{2016}  \at{Turbulent magnetohydrodynamic reconnection mediated by the plasmoid instability}.  \jt{The Astrophysical Journal}  \bvol{818}~(1),  \pg{20}.

\bibitem[Hudson {\em et~al.\/}(2012)Hudson, Dewar, Dennis, Hole, McGann, von Nessi \& Lazerson]{Hudson2012}
{\sc \au{Hudson, S.~R.}, \au{Dewar, R.~L.}, \au{Dennis, G.}, \au{Hole, M.~J.}, \au{McGann, M.}, \au{von Nessi, G.} \& \au{Lazerson, S.}} \yr{2012}  \at{Computation of multi-region relaxed magnetohydrodynamic equilibria}.  \jt{Physics of Plasmas}  \bvol{19}.

\bibitem[Kong(2020)]{Kong2020}
{\sc \au{Kong, Mengdi}} \yr{2020}  \at{Towards integrated control of tokamak plasmas: physics-based control of neoclassical tearing modes in the tcv tokamak}  \pg{p. 228}.

\bibitem[Kong {\em et~al.\/}(2022)Kong, Felici, Sauter, Galperti, Vu, Ham, Hender, Maraschek, Reich, the TCV~team, the ASDEX Upgrade~team, the MAST~team \& the EUROfusion MST1~Team]{Kong_2022}
{\sc \au{Kong, M}, \au{Felici, F}, \au{Sauter, O}, \au{Galperti, C}, \au{Vu, T}, \au{Ham, C~J}, \au{Hender, T~C}, \au{Maraschek, M}, \au{Reich, M}, \au{the TCV~team}, \au{the ASDEX Upgrade~team}, \au{the MAST~team} \& \au{the EUROfusion MST1~Team}} \yr{2022}  \at{Physics-based control of neoclassical tearing modes on tcv}.  \jt{Plasma Physics and Controlled Fusion}  \bvol{64}~(4),  \pg{044008}.

\bibitem[Kumar(2022)]{Kumar_2022}
{\sc \au{Kumar, A.}} \at{ \yr{2022} } \jt{Plasma Physics and Controlled Fusion}  \bvol{64}~(6),  \pg{065001}.

\bibitem[Loizu \& Bonfiglio(2023)]{Loizu_Bonfiglio_2023}
{\sc \au{Loizu, J.} \& \au{Bonfiglio, D.}} \yr{2023}  \at{Nonlinear saturation of resistive tearing modes in a cylindrical tokamak with and without solving the dynamics}.  \jt{Journal of Plasma Physics}  \bvol{89}~(5),  \pg{905890507}.

\bibitem[Loizu {\em et~al.\/}(2020)Loizu, Huang, Hudson, Baillod, Kumar \& Qu]{Loizu2020}
{\sc \au{Loizu, J.}, \au{Huang, Y.~M.}, \au{Hudson, S.~R.}, \au{Baillod, A.}, \au{Kumar, A.} \& \au{Qu, Z.~S.}} \yr{2020}  \at{Direct prediction of nonlinear tearing mode saturation using a variational principle}.  \jt{Physics of Plasmas}  \bvol{27}.

\bibitem[Loureiro {\em et~al.\/}(2005)Loureiro, Cowley, Dorland, Haines \& Schekochihin]{Loureiro2005}
{\sc \au{Loureiro, N.~F.}, \au{Cowley, S.~C.}, \au{Dorland, W.~D.}, \au{Haines, M.~G.} \& \au{Schekochihin, A.~A.}} \yr{2005}  \at{$x$-point collapse and saturation in the nonlinear tearing mode reconnection}.  \jt{Phys. Rev. Lett.}  \bvol{95},  \pg{235003}.

\bibitem[Militello \& Porcelli(2004)]{Militello2004}
{\sc \au{Militello, Fulvio} \& \au{Porcelli, F.}} \yr{2004}  \at{Simple analysis of the nonlinear saturation of the tearing mode}.  \jt{Physics of Plasmas}  \bvol{11},  \pg{L13}.

\bibitem[Muraglia {\em et~al.\/}(2021)Muraglia, Poyé, Agullo, Dubuit \& Garbet]{Muraglia2021}
{\sc \au{Muraglia, M.}, \au{Poyé, A.}, \au{Agullo, O.}, \au{Dubuit, N.} \& \au{Garbet, X.}} \yr{2021}  \at{Nonlinear dynamics of ntm seeding by turbulence}.  \jt{Plasma Physics and Controlled Fusion}  \bvol{63}.

\bibitem[Rutherford(1973)]{Rutherford1973}
{\sc \au{Rutherford, P.~H.}} \yr{1973}  \at{{Nonlinear growth of the tearing mode}}.  \jt{The Physics of Fluids}  \bvol{16}~(11),  \pg{1903--1908}.

\bibitem[Sauter {\em et~al.\/}(2002)Sauter, Buttery, Felton, Hender, Howell \& contributors to~the EFDA-JET~Workprogramme]{OSauter_2002}
{\sc \au{Sauter, O}, \au{Buttery, R~J}, \au{Felton, R}, \au{Hender, T~C}, \au{Howell, D~F} \& \au{contributors to~the EFDA-JET~Workprogramme}} \yr{2002}  \at{Marginal beta-limit for neoclassical tearing modes in jet h-mode discharges}.  \jt{Plasma Physics and Controlled Fusion}  \bvol{44}~(9),  \pg{1999}.

\bibitem[Taylor(1974)]{Taylor1974}
{\sc \au{Taylor, J.~B.}} \yr{1974}  \at{Relaxation of toroidal plasma and generation of reverse magnetic fields}.  \jt{Phys. Rev. Lett.}  \bvol{33},  \pg{1139--1141}.

\bibitem[Teng {\em et~al.\/}(2018)Teng, Ferraro, Gates \& White]{Teng_2018}
{\sc \au{Teng, Q.}, \au{Ferraro, N.}, \au{Gates, D.A.} \& \au{White, R.B.}} \yr{2018}  \at{Nonlinear simulations of thermo-resistive tearing mode formalism of the density limit}.  \jt{Nuclear Fusion}  \bvol{58}~(10),  \pg{106024}.

\bibitem[Watanabe {\em et~al.\/}(2005)Watanabe, Sakakibara, Narushima, Funaba, Narihara, Tanaka, Yamaguchi, Toi, Ohdachi, Kaneko, Yamada, Suzuki, Cooper, Murakami, Nakajima, Yamada, Kawahata, Tokuzawa, Komori \& experimental group]{Watanabe_2005}
{\sc \au{Watanabe, K.Y.}, \au{Sakakibara, S.}, \au{Narushima, Y.}, \au{Funaba, H.}, \au{Narihara, K.}, \au{Tanaka, K.}, \au{Yamaguchi, T.}, \au{Toi, K.}, \au{Ohdachi, S.}, \au{Kaneko, O.}, \au{Yamada, H.}, \au{Suzuki, Y.}, \au{Cooper, W.A.}, \au{Murakami, S.}, \au{Nakajima, N.}, \au{Yamada, I.}, \au{Kawahata, K.}, \au{Tokuzawa, T.}, \au{Komori, A.} \& \au{experimental group, LHD}} \yr{2005}  \at{Effects of global mhd instability on operational high beta-regime in lhd}.  \jt{Nuclear Fusion}  \bvol{45}~(11),  \pg{1247}.

\end{thebibliography}

\end{document}